%% file: main.tex
\documentclass[sigconf]{acmart}

\usepackage{caption}
\usepackage{subcaption}

\usepackage{epsfig}
\usepackage{graphicx}
\usepackage{amsmath}

\usepackage{amssymb}
\usepackage[utf8]{inputenc} 
\usepackage[T1]{fontenc}    
\usepackage{xurl}            
\usepackage{booktabs}       
\usepackage{amsfonts}       
\usepackage{nicefrac}       
\usepackage{microtype}      
\usepackage{xcolor}         
\usepackage{microtype}
\usepackage{graphicx}
\usepackage{relsize}
\usepackage{tabularx}
\usepackage{booktabs} 
\usepackage{sidecap}
\sidecaptionvpos{figure}{c}
\usepackage{algorithm}
\usepackage[noend]{algpseudocode}
\usepackage{enumitem}
\newlist{steps}{enumerate}{1}
\setlist[steps, 1]{label = Step \arabic*:}
\usepackage{bm}
\usepackage{comment}
\usepackage{xspace}
\usepackage{wrapfig,lipsum,booktabs}
\usepackage{tikz}
\usepackage{framed}
\usepackage{pifont}
\usepackage{tablefootnote}
\usepackage{mathtools}
\usepackage{amsthm}

\setcitestyle{numbers}

\newif\ifdraft
\draftfalse

\ifdraft
\newcommand{\nojan}[1]{\textcolor{red}{{\sf (NS:} {\sl{#1})}}}
\newcommand{\anees}[1]{\textcolor{blue}{{\sf (AA:} {\sl{#1})}}}
\newcommand{\nges}[1]{\textcolor{orange}{{\sf (Ng:} {\sl{#1})}}}
\newcommand{\davi}[1]{\textcolor{purple}{{\sf (DA:} {\sl{#1})}}}

\else
\newcommand{\nojan}[1]{}
\newcommand{\anees}[1]{}
\newcommand{\nges}[1]{}
\newcommand{\davi}[1]{}
\fi

\newcommand{\Prv}{$\mathcal{P}$\xspace}
\newcommand{\Vrf}{$\mathcal{V}$\xspace}
\newcommand{\Cir}{$\mathcal{C}$\xspace}
\newcommand{\sys}{\texttt{AMAZE}\xspace}

\algblockdefx[SWITCH]{Switch}{EndSwitch}[1]{\textbf{switch} #1 \textbf{when}}{\textbf{end} \textbf{switch}}%
\algblockdefx[CASE]{Case}{EndCase}[1]{\textbf{case} #1 \textbf{do}}{\textbf{end} \textbf{case}}%
\algblockdefx[FIXEDLOOP]{FixedLoop}{EndFixedLoop}[1]{\textbf{loop} #1 \textbf{iterations}}{\textbf{end} \textbf{loop}}%

\algrenewcommand\algorithmicrequire{\textbf{Input:}}
\algrenewcommand\algorithmicensure{\textbf{Output:}}

\newcommand{\leftshift}{<\hspace{-0.2em}<}
\newcommand{\rightshift}{>\hspace{-0.2em}>}

\AtBeginDocument{%
  }

\copyrightyear{2024}
\acmYear{2024}
\setcopyright{acmlicensed}
\acmConference[ICCAD '24]{IEEE/ACM International Conference on Computer-Aided Design}{October 27-October 31, 2024}{Newark, NJ, USA}
\acmBooktitle{IEEE/ACM International Conference on Computer-Aided Design (ICCAD '24), October 27-October 31, 2024, Newark, NJ, USA}
\acmDOI{10.1145/3676536.3676809}
\acmISBN{979-8-4007-1077-3/24/10}

\begin{document}

\title{AMAZE: \underline{A}ccelerated \underline{M}iMC Hardware \underline{A}rchitecture for \underline{Z}ero-Knowledge Applications on the \underline{E}dge} 

\author{Anees Ahmed$^{1}$\text{*}, Nojan Sheybani$^{2}$\text{*}, Davi Moreno$^{1}$, Nges Brian Njungle$^1$, Tengkai Gong$^2$, Michel~Kinsy{$^1$},~Farinaz Koushanfar{$^2$}}
\affiliation{
\department{\text{*}Equal contribution}
  \institution{$^1$Arizona State University, $^2$University of California San Diego}
    $^1$\tt\small{\{aahmed90, dcdealme, nnjungle, mkinsy\}@asu.edu}, $^2$\tt\small{\{nsheyban, tegong, farinaz\}@ucsd.edu}
    \country{}\\
 }

\renewcommand{\shortauthors}{Ahmed \& Sheybani, et al.}

\input{sections/0_abstract}



\received{5 May 2024}
\received[revised]{5 May 2024}
\received[accepted]{30 June 2024}

\maketitle

\input{sections/1_intro}
\input{sections/2_prelim}
\input{sections/3_related}
\input{sections/4_methodology}
\input{sections/5_results}
\input{sections/6_conclusion}

\section{Acknowledgements}
This work was supported by DARPA Proofs under grant number HR0011-23-1-0006. The authors would like to thank Dr. Bryant York for his tireless guidance and mentorship throughout this project.
\vspace{-2mm}

\bibliographystyle{ACM-Reference-Format}
\bibliography{main}

\end{document}

%% file: sections/0_abstract.tex
\begin{abstract}
  Collision-resistant, cryptographic hash (CRH) functions have long been an integral part of providing security and privacy in modern systems. Certain constructions of zero-knowledge proof (ZKP) protocols aim to utilize CRH functions to perform cryptographic hashing. Standard CRH functions, such as SHA2, are inefficient when employed in the ZKP domain, thus calling for \textit{ZK-friendly hashes}, which are CRH functions built with ZKP efficiency in mind. The most mature ZK-friendly hash, MiMC, presents a block cipher and hash function with a simple algebraic structure that is well-suited, due to its achieved security and low complexity, for ZKP applications. Although ZK-friendly hashes have improved the performance of ZKP generation in software, the underlying computation of ZKPs, including CRH functions, must be optimized on hardware to enable practical applications. The challenge we address in this work is determining how to efficiently incorporate ZK-friendly hash functions, such as MiMC, into hardware accelerators, thus enabling more practical applications. In this work, we introduce \texttt{AMAZE}, a highly hardware-optimized open-source framework for computing the MiMC block cipher and hash function. Our solution has been primarily directed at resource-constrained edge devices; consequently, we provide several implementations of MiMC with varying power, resource, and latency profiles. Our extensive evaluations show that the \texttt{AMAZE}-powered implementation of MiMC outperforms standard CPU implementations by more than 13$\times$. In all settings, \texttt{AMAZE} enables efficient ZK-friendly hashing on resource-constrained devices. Finally, we highlight \texttt{AMAZE}'s underlying open-source arithmetic backend as part of our end-to-end design, thus allowing developers to utilize the \texttt{AMAZE} framework for custom ZKP applications.
\end{abstract}

%% file: sections/1_intro.tex
\section{Introduction}

As data privacy and security have been more of a concern in the past decade, the concept of privacy-preserving computation, which enables computation to be performed on encrypted data, has been introduced as a paradigm shift in computing. Specifically, zero-knowledge proofs (ZKPs), a privacy-preserving cryptographic primitive, allow users to prove certain attributes about their private data without revealing anything about the data. Although ZKPs in their current state of implementation on software have proven to be effective in many applications, such as authentication \cite{lu2008pseudo, liu2011zero}, healthcare \cite{sharma2020blockchain, gaba2022zero}, and emerging learning paradigms \cite{ghodsi2023zprobe, liu2021zkcnn, weng2021mystique}, designing ZKP applications requires careful software/algorithm co-design to ensure that computation achieves practical runtimes and resource utilization on modern systems. As computational overhead is the main challenge when building practical zero-knowledge systems, there has been a recent emergence of research and development on ZKP hardware accelerators \cite{ma2023gzkp, zhang2021pipezk}. While these accelerators have proven to be effective, they often target FPGA or ASIC devices with high computational power or a large amount of resources.

ZKPs have been shown to be a very valuable primitive in the IoT \cite{chen2023survey, boo2021litezkp} and other edge computing workflows \cite{wu2020blockchain}, which often perform computation on resource-constrained devices. These use cases further motivate the importance of ZKP hardware acceleration while introducing a novel challenge: catering custom ZKP hardware to resource-constrained edge devices. Before an end-to-end accelerator can be built to maximize efficiency with limited resources, certain underlying modules must be built and highly optimized for area and latency. In this work, we focus on two core computational building blocks for the seminal ZKP constructions - collision-resistant, cryptographic hash functions and Galois/finite field arithmetic.

Collision-resistant, cryptographic hash (CRH) functions have served as a powerful tool to enable secure computation and storage in modern systems. Certain constructions of zero-knowledge proof (ZKP) protocols, such as zk-STARKs \cite{ben2018scalable} and zk-SNARKs \cite{nitulescu2020zk}, utilize CRH functions to perform cryptographic hashing for various applications. Traditional NIST-approved CRH functions, such as SHA-2 and SHA-3, have been proven to be secure through extensive studies and applications. There have been comprehensive efforts towards the thorough design of hardware and software to ensure their efficiency \cite{sklavos2003hardware}. However, efficiency in the ZKP domain is dependent on several factors that are not accounted for in our current systems, such as algebraic structure and multiplicative complexity. These standard CRH functions have proven to be inefficient when translated to the ZKP domain, thus calling for \textit{ZK-friendly hashes}, which are CRH functions built with ZKP efficiency in mind \cite{Ingonyama_2022}. In this work, we consider MiMC, the first and most mature ZK-friendly hash \cite{albrecht2016mimc}, consisting of a block cipher and hash function with a simplified algebraic structure that was originally designed for zk-SNARKs, but has found use in zk-STARK applications as well \cite{ben2020stark}. While the algebraic structure of MiMC is relatively simple, the underlying arithmetic structure is a large Galois prime field, which typically requires computation to be done on 254-bit integers. We do note that the size of the prime field is dependent on the ZKP construction, but typically nothing less than 128 bits would be used for security purposes. Nonetheless, efficient prime field arithmetic is a very challenging task to do on resource-constrained devices, such as select FPGAs, as the hardware modules for performing fast arithmetic typically only support 16 to 27-bit arithmetic.
To address this, we present \texttt{AMAZE}, an accelerated hardware architecture that enables underlying fast and resource-efficient Galois field arithmetic for the MiMC hash function on resource-constrained edge devices. This work provides an accessible solution for developers and businesses to incorporate the MiMC hash function on low-end FPGAs for custom zero-knowledge applications.

In short, our contributions are as follows:
\begin{itemize}
    \item We propose \texttt{AMAZE}, a highly-optimized hardware architecture framework for computing the MiMC block cipher and hash function, a core operation in zero-knowledge proofs, on FPGA. \texttt{AMAZE} is designed to support resource-constrained edge devices, while still outperforming CPU.
    \item Our open-source implementation\footnote{\url{https://github.com/ACES-STAM/AMAZE}} is parameterizable to balance power, resource utilization, and latency based on the available resources, without sacrificing the security of the MiMC hash function. This is done through our novel design of the well-established Russian Peasant and Barrett modular multiplication schemes. We provide an open-source implementation of optimized Galois field arithmetic library that is compatible with the BN254 elliptic curve, a commonly used elliptic curve in zk-SNARKs.
    \item Our extensive evaluations show that our novel, fully pipelined implementation, which uses Barrett Reduction for modular multiplication, achieves more than 13$\times$ speedup (for one block cipher invocation or one hash round) when compared to state-of-the-art MiMC software running on a server-grade CPU. This performance achieves relatively low power consumption on a low-end FPGA with limited resources, highlighting the feasibility of \texttt{AMAZE} on resource-constrained edge devices for zero-knowledge proof applications.
\end{itemize}

%% file: sections/2_prelim.tex
\section{Preliminaries}

\subsection{Zero-Knowledge Proofs}
Zero-Knowledge Proofs (ZKPs) are a cryptographic primitive in which a prover \Prv proves to a verifier \Vrf that they know a secret value $w$, referred to as the witness, without revealing anything about $w$. Computation in the ZK domain is often expressed as a circuit \Cir, which can informally be thought of as a function that takes in public and private inputs, and generates a public output. Formally, \Prv generates a proof attesting that they know a secret value $w$ such that $C(x; w) = y$, in which $x$ and $y$ are public inputs and outputs, respectively. Note that such a proof does not reveal anything about the witness $w$ to anybody, including the verifier \Vrf. Before a proof can be generated in ZKP schemes, the circuit \Cir must go through a process called \textit{arithmetization}, in which the computation is represented in efficient mathematical terms (e.g. polynomials). For brevity's sake, we treat arithmetization as a black-box, in which \Cir is converted into a ZKP scheme-specific mathematical representation. Below, we briefly outline zk-SNARKs and zk-STARKs, two of the seminal ZKP constructions.

\textbf{Zero-Knowledge Succinct Non-Interactive Arguments of Knowledge (zk-SNARKs)} are a class of non-interactive ZKP protocols that guarantee succinct proof size. The most commonly used proving scheme in zk-SNARK constructions is the Groth16 protocol~\cite{groth2016size}, which generates publicly-verifiable proofs, around 128 bytes.
The underlying cryptography for Groth16 is computationally intensive elliptic curve cryptography (ECC) \cite{ben2017scalable}.
The main hindrance of using Groth16 zk-SNARKs in practical settings is its reliance on a computationally heavy trusted setup process per circuit \Cir, which is necessary to generate protocol keys for \Prv and \Vrf to assist with proof generation and verification. This, however, is often acceptable in practice in scenarios where \Cir is not changing.
 
\textbf{Zero-Knowledge Scalable Transparent Arguments of Knowledge (zk-STARKs)} are a class of ZKP protocols that remove the dependence on the trusted setup process. 
Rather than using ECC, zk-STARKs utilize lightweight, post-quantum safe cryptography for proof generation, relying on collision-resistant hash functions. These hash functions lend themselves nicely to proof generation, as the main underlying data structure is the Merkle tree, which often results in non-succinct proofs \cite{ben2018scalable}.
Although not as efficient as zk-SNARK verification, zk-STARK verification is kept somewhat efficient by only verifying certain critical paths of the Merkle tree, which are described in the generated proof, instead of the whole Merkle tree.

\subsection{ZK-Friendly Hash Functions}
One \textit{general} metric that is used to measure the size of ZK arithmetic representations is the number of constraints. In brief, constraints are an intermediate form of representation of the circuit \Cir that helps form the final mathematical statement, but these constraints generally indicate how complex a circuit is in the ZK domain. The goal of ZK-friendly hashes is to perform cryptographically secure collision-resistant hashing, while only requiring a minimal amount of constraints. Hash functions are very often used in zk-SNARK and zk-STARK prover and verifier algorithms for hash chaining, Merkle trees, commitments, and integrity checks with proofs of hash preimages. 
Our work focuses on hardware acceleration of MiMC, a ZK-friendly hash function which is optimized to minimize the amount of constraints in zk-SNARKs, but has proven to be effective in zk-STARKs too \cite{ben2020stark}. 
There has been a notable increase in research that presents new ZK-friendly hash functions, such as Vision/Rescue \cite{aly2020design}, GMiMC \cite{albrecht2019feistel}, and Poseidon \cite{grassi2021poseidon}. While each hash function aims to optimize for different parameters, the overarching goal of each work is to produce a low-overhead approach for secure hashing in the ZK domain. We prioritize MiMC in this work due to its prominence in current ZKP-based systems \cite{khovratovich2019tornado, chen2022review}, but note that the underlying modules that we present in \sys can be utilized to accelerate other notable ZK-friendly hash functions as well, especially those that rely on Galois field arithmetic. 

\subsection{Galois Field Arithmetic}
Galois field arithmetic, commonly referred to as finite field arithmetic, ensures that the inputs and outputs of all arithmetic operations are contained within a finite mathematical field \cite{benvenuto2012galois}. The size of an integers-based Galois field in cryptographic settings is represented by a large prime number $p$. Such fields are denoted as GF($p$) and are often referred to as \textit{prime fields}.
Arithmetic within a field GF($p$) is relatively straightforward --- all operations, such as multiplication and addition, are performed modulo $p$ to ensure that there are no values greater than $p-1$ or lower than $0$. Most hardware architectures are capable of performing Galois Field arithmetic efficiently, however, in the cryptographic setting where $p$ is a very large number, modular multiplication becomes a challenging task to optimize. In zk-STARKs, the field that all arithmetic is performed in can be chosen for performance, but in zk-SNARKs, the field must be chosen as a prime scalar subfield of a pairing-friendly elliptic curve. Such elliptic curves have quite a large subfield in order to ensure sufficient security level. BN-254, a widely-used elliptic curve in zk-SNARKs, requires $p$ in GF($p$) to be $254$ bits wide.

As these large integers are inefficient to multiply and reduce modulo $p$ on hardware if modular reduction is naively done using long division, we require the use of efficient algorithms for modular multiplication. In this work, we utilize two well-known modular $n$-bit multiplication methods: the Russian peasant method~\cite{pan2013bit}, where the number of clock cycles is equal to the number of bits in multiplicands, and the Barrett reduction method, which uses a sequence of 3 integer multiplications and two integer subtractions~\cite{barrett1986implementing}. 
The Russian Peasant method, shown in Algorithm~\ref{alg:GFp_mult_peasant}, proceeds bit-by-bit, using only addition and bit-shift operations. Note that in this listing, $\Call{LSB}{x}$ denotes the least significant bit of $x$, and $\leftshift$ and $\rightshift$ denote the left and right bit-shift operators respectively. Although this approach is slow, it greatly reduces the amount of necessary resources and does not require the use of DSPs at all. Comparatively, the Barrett method, shown in Algorithm~\ref{alg:GFp_mult_barrett}, is more resource intensive, but is also one of the fastest ways to compute modular multiplication on hardware \cite{langhammer2021efficient}. While Montgomery reduction \cite{koc1996analyzing} is also similarly fast, it requires that integers to be converted in and out of ``Montgomery'' form, which is an expensive operation, so we do not consider this reduction scheme in this work. 

\begin{algorithm}[htb]
    \caption{Russian Peasant Method for Modular Multiplication}
    \label{alg:GFp_mult_peasant}
    \begin{algorithmic}[1]
        \Require $x_1, x_2 \in$ GF($p$) \enspace such that:
        \Statex $n = \lceil \log_2 (p-1) \rceil$
        \Ensure $y = (x_1 \times x_2) \bmod p$
        \State $y \gets 0$
        \FixedLoop{$n$}
            \State $t \gets$ \textbf{if} $\Call{LSB}{x_2} = 1$ \textbf{then} $x_1$ \textbf{else} $0$
            \State $y \gets y + t$ 
            \State $y \gets$ \textbf{if} $y > p$ \textbf{then} $y - p$ \textbf{else} $y$ 
            \State $u \gets x_1 \mathbin{\leftshift} 1$ \Comment{Multiplying $x_1$ by 2}
            \State $x_1 \gets$ \textbf{if} $u > p$ \textbf{then} $u - p$ \textbf{else} $u$
            \State $x_2 \gets x_2 \mathbin{\rightshift} 1$ \Comment{Dividing $x_2$ by 2}
        \EndFixedLoop
    \end{algorithmic}
\end{algorithm}

\begin{algorithm}[htb]
    \caption{Barrett Method for Modular Multiplication}
    \label{alg:GFp_mult_barrett}
    \begin{algorithmic}[1]
        \Require $x_1, x_2 \in$ GF($p$) \enspace \textbf{and} \enspace $z$, $n$ \enspace such that: 
        \Statex $n = \lceil \log_2(p-1) \rceil$
        \Statex $z = \lfloor 2^{2n} \mathbin{/} p \rfloor$
        \Ensure $y = (x_1 \times x_2) \bmod p$
        \State $w \gets x_1 \times x_2$
        \State $t \gets \big( w \mathbin{\rightshift} (n - 1) \big) \times z$
        \State $u \gets \big( t \mathbin{\rightshift} (n + 1) \big) \times p$ 
        \State $y \gets \big( w \bmod 2^{n + 1} \big) - \big( u \bmod 2^{n + 1} \big)$ 
        \State $y \gets$ \textbf{if} $y > p$ \textbf{then} $y - p$ \textbf{else} $y$
        \State $y \gets$ \textbf{if} $y > p$ \textbf{then} $y - p$ \textbf{else} $y$
    \end{algorithmic}
\end{algorithm}

\subsection{MiMC}
MiMC is a block cipher and hash function family designed with low multiplicative complexity. The MiMC hash function was designed to maximize zk-SNARK friendliness, but it also achieved very low complexity for zk-STARKs. As the most mature ZK-friendly hash function, MiMC has been implemented in many real-world systems that utilize ZKPs and hash functions \cite{khovratovich2019tornado, chen2022review}.
The original design of the MiMC block cipher is simple --- the core computation is $f(x)=x^3$ in the chosen binary field GF($2^n$), which is repeated multiple times and incorporates round constants and keys. However, MiMC can be generalized to work over prime fields as well. The core computation can be $f(x)=x^d$ for any integer $d$, as long as gcd$(d, p-1)=1$ for the chosen field GF($p$). We will discuss this in depth. The full block cipher computation is carried out by performing $r$ rounds, each round function $F_i$ being 
\begin{equation*}
    F_{0 \leq i \leq r}(x) = (x\mathbin{\oplus}k\mathbin{\oplus}c_i)^d
\end{equation*}
\noindent where $x$ is the initial input to the first round ($i=0$), or the output of the previous round ($i>0$), $k$ is a key, and $c_i$ is a precomputed round constant. In this paper, $\oplus$ denotes addition in a Galois field. The round constants must be publicly agreed upon in order for computed hashes to be compatible.
We refer to the cipher construction as MiMC-$p/p$ and MiMC-$n/n$ where it operates over the fields GF($p$) and GF($2^n$) respectively. The required number of rounds is computed as $r=\left\lceil log_2(N) \mathbin{/} log_2(d)\right\rceil$ where $N$ is the size of the field \cite{albrecht2016mimc}. 
For the rest of the paper, we will focus on the variant MiMC-$p/p$, as our work performs computation over a prime Galois field. In the ZK domain, this computation is typically done over prime scalar subfields of pairing-friendly elliptic curves, such as BN-254 \cite{barreto2005pairing}, BLS12-377 \cite{bowe2020zexe}, and BLS12-381 \cite{barreto2003constructing}.

\begin{figure}
    \centering
    \includegraphics[width=0.8\columnwidth]{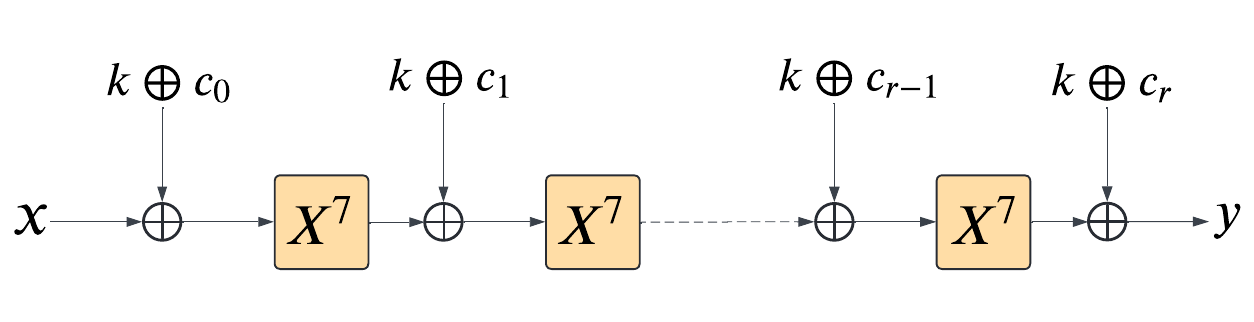}
    \caption{The MiMC-$p/p$ block cipher that we accelerate on FPGA. Please note that round constants $c_i$ are precomputed and key $k$ is user-supplied. In our work, number of rounds $r=91$ whenever we work with BN-254.}
    \label{fig:mimc-cipher}
\end{figure}

In this work, we use the scalar prime subfield of the BN-254 elliptic curve. Therefore, we choose $d=7$ and modify the round function accordingly. We calculate the number of rounds to be $r=91$. The MiMC-$p/p$ block cipher is denoted as MiMC($x$, $k$) = $y$, where $x$ is a message to encrypt, $k$ is an encryption key, and $y$ is an encrypted output. Figure~\ref{fig:mimc-cipher} illustrates the MiMC-$p/p$ block cipher.

\begin{figure}
    \centering
    \includegraphics[width=0.5\columnwidth]{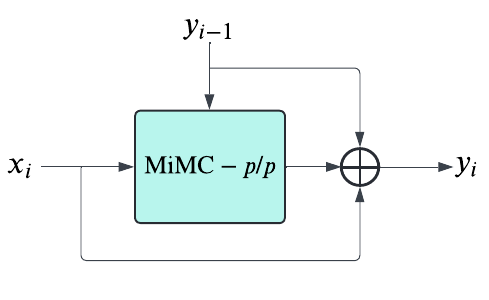}
    \caption{The MiMC-$p/p$-based hash function that is accelerated on FPGA by \sys.}
    \label{fig:mimc-hash}
\end{figure}

Finally, a MiMC-$p/p$-based hash function can be constructed by using the block cipher as its underlying security mechanism. While there are several ways to construct the hash function, such as the Merkle--Damg{\aa}rd \cite{coron2005merkle} and sponge constructions \cite{bertoni2007sponge}, we opt for the Miyaguchi--Preneel \cite{menezes2018handbook} hash construction, as it is prevalent in real-world applications, such as Ethereum \cite{githubGitHubHarryRethsnarks}. For padding, any Merkle-Damg{\aa}rd-like message padding scheme can be constructed. This is also a relatively simple construction that can be represented easily in the ZK domain. This hash function consists of multiple rounds: one round for each message block. The $i^{\text{th}}$ round of this hash function is expressed as
\begin{equation*}
    y_i = \text{MiMC}(y_{i-1}, x_i) \mathbin{\oplus} y_{i-1} \mathbin{\oplus} x_i
\end{equation*}
\noindent where the MiMC($y_{i-1}$, $x_i$) block cipher function takes in $y_{i-1}$, the output of the previous hash round, as the key to the MiMC-$p/p$ cipher, and takes $x_i$, the current data block, as the message input. 
We show the MiMC-$p/p$-based hash function in Figure \ref{fig:mimc-hash}.

%% file: sections/3_related.tex
\section{Related Works}
While \sys is the first work to enable the MiMC block cipher and hash function on FPGA and specifically resource-constrained edge devices, there have been some notable hardware implementations of zero-knowledge proofs and ZK-friendly hash functions on more powerful hardware. PipeZK \cite{zhang2021pipezk} is the first notable work to propose hardware acceleration of ZKPs by introducing a hardware-optimized version of multiscalar multiplication (MSM) and number theoretic transform (NTT), the most computationally-intensive operations in zk-SNARKs. While effective, this work targets an ASIC implementation, making it a relatively inaccessible option compared to FPGA or GPU implementations. PipeMSM \cite{xavier2022pipemsm} proposes a novel methodology for MSM on FPGA, using an underlying optimized Barrett modular multiplication unit. This work achieves low latency with Barrett modular multiplication using optimizations that require the resources of a high-end board, namely the Xilinx Alveo U55C, which falls out of the realm of edge devices. Other efforts, such as \cite{githubGitHubZcashFoundationzcashfpga} and \cite{githubGitHubBsdevlinfpga_snark_prover}, which are open-source FPGA implementations of zk-SNARKs for Zcash and the Ethereum environment respectively, compute Montgomery modular multiplication. While these implementations are efficient when tested on the powerful Amazon F1 FPGA systems \cite{amazonAmazonInstances}, we also note that the boards that Amazon F1 provides are high-speed Xilinx Ultrascale+ devices that are powerful enough to accelerate deep neural networks. Although this work performs well on these powerful devices, the author theorized that the work would be at least 20\% faster if Barrett modular multiplication was used \cite{Bsdevlin}.
GPU acceleration of ZKPs has also been proposed \cite{ma2023gzkp, ni2023enabling}, but has proven to be an energy-intensive approach when compared to FPGAs. Thus, we do not consider GPU acceleration in our work due to our focus on resource-constrained edge devices.

In the realm of ZK-friendly hash function acceleration, TRIDENT \cite{githubGitHubDatenlordTRIDENT} accelerates the Poseidon hash function, another commonly used ZK-friendly hash, on a Xilinx Varium C1100 blockchain-friendly board. This work provides an excellent open-source implementation of the Poseidon hash function, however, the results show that there is still a high power and resource requirement. 
Alongside this, Poseidon, and its recent successor Poseidon2 \cite{grassi2023poseidon2}, have not been through as rigorous of a security analysis as MiMC has in practice, due to their relative nascence \cite{ben2020stark}.
As we've shown, there is a demand for accelerating ZKPs at all levels, and a common approach is to start with individual operations and build upwards (e.g. Galois Field arithmetic $\rightarrow$ hashing). The main challenge in our work is optimizing latency and resource utilization for ZK-friendly hashes on resource-constrained devices, while previous works have focused primarily on reducing latency at the cost of power and resources. 
We consider these works as adjacent to ours and believe \sys's optimizations can be applied to prior works for deployment in resource-constrained settings.

%% file: sections/4_methodology.tex
\section{Methodology}

\begin{figure}
    \centering
    \includegraphics[width=.7\columnwidth]{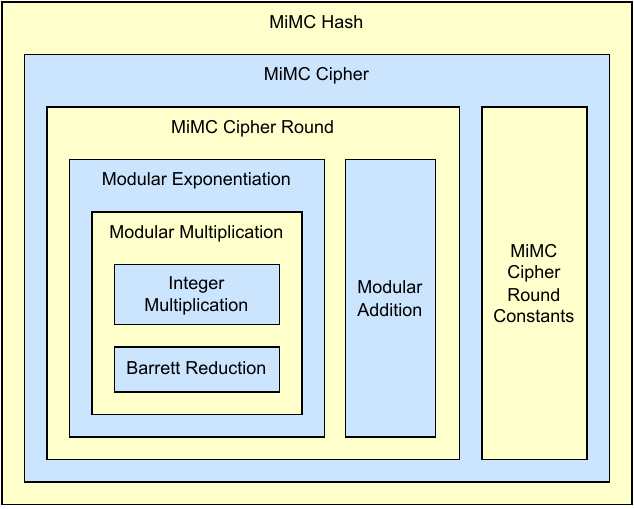}
    \caption{The hierarchical architecture of the \sys-powered MiMC accelerator.}
    \label{fig:mimc-high-level}
\end{figure}

The hierarchical architecture of the \sys-powered MiMC accelerator is depicted in Figure \ref{fig:mimc-high-level}. The explored optimizations, focused on timing or resource utilization, mainly involve the MiMC cipher round and modular multiplication modules. These optimizations include a fast 254-bit integer multiplier, a pipelined modular multiplier based on the Barrett reduction method, and a refined $x^7$ modular exponentiation unit, needed during each MiMC cipher round. We utilize these efficient modules to build an end-to-end pipelined hardware accelerator for the MiMC hash function on FPGA with state-of-the-art performance.
In the subsequent subsections, we highlight our optimization techniques aimed at addressing these computational bottlenecks and our final, most-efficient design.

In every invocation of the MiMC block cipher, there are 91 rounds. The time spent in each round is dominated by the exponentiation operation, as the latency of adding MiMC round constants and the key is relatively small. The exponentiation operation (to the $7^{\text{th}}$ power) requires four large modular multiplications. Each such modular multiplication requires three large integer multiplications, as we use the Barrett reduction method. In total, the whole block cipher computation is dominated by $91 \times 4 \times 3$ large integer multiplications. To complete the block cipher computation in minimum time, the hardware design must focus on completing these $91 \times 4 \times 3$ multiplications as fast as possible. Note that neither the cipher rounds nor the hash rounds are parallelizable, as the input of any round depends on the output of the previous round. Therefore, the MiMC block cipher and the Miyaguchi--Preneel hash function are both very \textit{serial} computations. However, the throughput can still be boosted by computing multiple (but independent) cipher or hash computations in parallel. This is of great utility during batch MiMC computation requests, such as in the case of computing a Merkle tree root. The best way to build this while keeping hardware cost low is to use a multi-stage pipelined design. Therefore, in \sys, we present an optimized pipeline that completes a batch of 13 block cipher computations in 4823 cycles, designed such that it can be run at a clock frequency of \textasciitilde 129 MHz. This particular design is referred to as \texttt{AMZ-1} in this paper.

\subsection{Notation} This sub-section establishes some of the general notations and assumptions used throughout this ``Methodology'' section. Consider an integer $x$ that is $n$ bits wide. Its least significant bit is at the index 0 and its most significant bit is at the index $n-1$. The notation $x[b:a]$ denotes the chunk of bits from the smaller index $a$ to the larger index $b$, including the bits at the indices $a$ and $b$. We refer to the bits at smaller indices as the \textit{lower} bits, and the ones at higher indices as the \textit{upper} bits.

\subsection{Optimized 254-bit integer multiplier}
\label{sec:int-multiplier}

When performing large integer multiplication in hardware, careful design is needed to adequately take advantage of the existing hardware resources. Most FPGA boards have built-in fast multiply and accumulate units, called DSPs, which can be used for rapid integer multiplication. Generally, DSPs support multiplication of small integers, such as 27-bit integer multiplication in the Stratix V GT FPGA~\cite{intelStratixFPGA}. When presented with a simple hardware description of multiplication, such as ``\texttt{out <= x*y;}'', most FPGA design tools will utilize DSPs to automatically implement the multiplication logic with the available resources. However, challenges arise when the involved integers have a bit-width that exceeds the DSPs' capabilities, necessitating optimizations by the design tools, which may result in inefficient designs with excessive use of DSPs, or other resources, or poor timing causing long clock cycles.

\begin{figure}
    \centering
    \includegraphics[width=0.8\columnwidth]{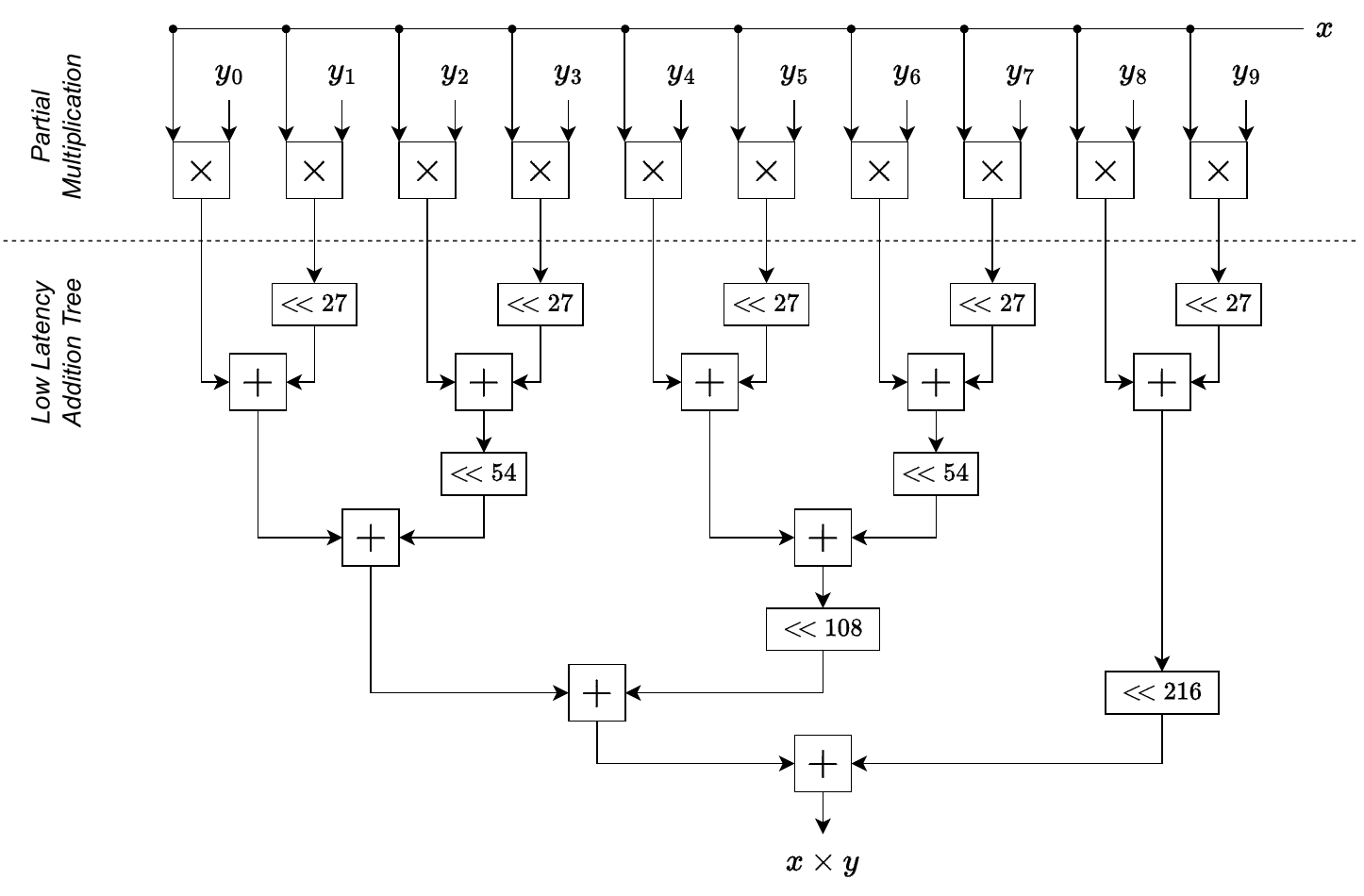}
    \caption{The design of the fast 254-bit integer multiplier. $x$ and $y$ are the multiplicands. $y_i$ denotes the 27-bit chunk $y[27(i+1)-1 : 27i]$. The ``$\times$'' blocks are 27-bit multipliers and the ``$+$'' blocks are adders. The ``$\leftshift$'' blocks represent bit-wise left shift operations.}
    \label{fig:mimc-mult}
\end{figure}

To overcome this problem, we split the multiplication into two distinct steps, as illustrated in Figure~\ref{fig:mimc-mult}. Consider the multiplication of two 254-bit integers $x$ and $y$. The second operand $y$ can be split into ten 27-bit chunks $y_0$, $y_1$, \ldots, $y_9$, where $y_0$ has the least significant bits. Mathematically:
\begin{equation*}
    y = \sum_{i} y_i \cdot 2^{27 i}
\end{equation*}
The first step, referred to as \textit{partial multiplication}, performs smaller multiplications of $x$ with individual 27-bit chunks $y_i$; we observed that this computation can be well optimized automatically by the design tools. The results of these small multiplications are stored in registers for the next step. The second step, referred to as \textit{low-latency addition tree}, adds all these partial products together using a sequence of large integer additions and bit-shift operations. This sum is the final multiplication result. This works because mathematically:
\begin{equation*}
    x \cdot y = x \cdot \sum_{i} y_i \cdot 2^{27 i} = \sum_{i} x \cdot y_i \cdot 2^{27 i}
\end{equation*}
Simply writing a hardware description that expresses a straightforward addition of all these small partial products is not desirable, because the design tools infer a circuit that has high latency due to long carry wires. The tree structure of the proposed addition circuit prevents the carry wires from being too long in the inferred circuit, and hence the complete tree circuit is able to fit within a single clock cycle period without decreasing the clock frequency.

\begin{figure}
    \centering
    \includegraphics[width=0.8\columnwidth]{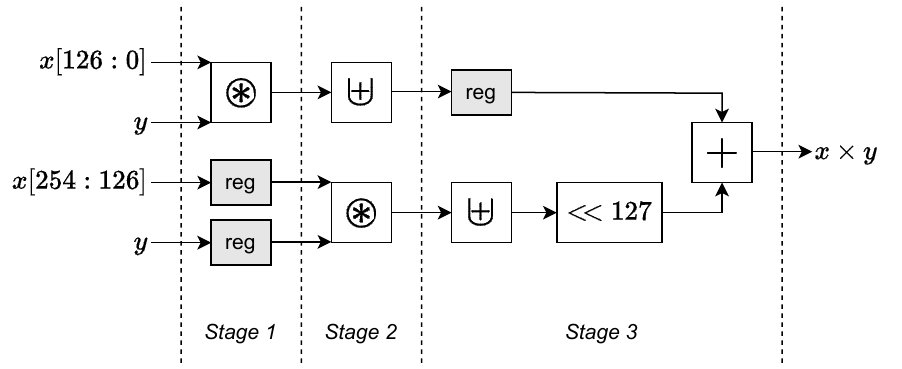}
    \caption{The design of the 3-stage pipeline for the fast 254-bit integer multiplier. $x$ and $y$ are the multiplicands. The ``$+$'' blocks are adders and the ``$\leftshift$'' blocks represent bit-wise left shift operations. The ``$\circledast$'' and ``$\uplus$'' blocks represent the ``partial multiplication'' and ``low-latency addition tree'' portions, respectively, of the multiplier hardware shown in Figure~\ref{fig:mimc-mult}. The ``reg'' blocks represent registers that simply hold data.}
    \label{fig:mimc-mult-pipeline}
    \vspace{-3mm}
\end{figure}

However, simply using this multiplication circuit to multiply two 254-bit integers does not necessarily give the best performance. Since this is still a large circuit, there is still room for improvement of the latency by decreasing its size. This can be achieved by splitting the first integer $x$ into multiple smaller parts, performing part-wise multiplications, and combining these results to get the final large product. In our experiments, we found that splitting $x$ into two parts achieves the best latency. Suppose that $x_0$ and $x_1$ are respectively the lower and upper halves of integer $x$. Then, mathematically:
\begin{equation*}
    x \cdot y = \left( \sum_{i} x_0 \cdot y_i \cdot 2^{27 i} \right) + \left( 2^{127} \sum_{i} x_1 \cdot y_i \cdot 2^{27 i} \right)
\end{equation*}
The primary advantage of utilizing these part-wise multiplications is unlocking the ability to perform them in two different clock cycles so that each clock cycle period can be relatively shorter. Thus, we propose the pipelined three-stage 254-bit multiplier shown in Figure~\ref{fig:mimc-mult-pipeline}. Each stage is completed in one clock cycle. The block with the ``$\times$'' symbol represents the first step (\textit{partial multiplication} and the one with the ``$+$'' symbol represents the second step (\textit{low-latency addition tree}), as described earlier. In stage 1, the partial products for $x_0 \cdot y$ are computed. In stage 2, the partial products for $x_1 \cdot y$ as well as the accumulated sum for $x_0 \cdot y$ are computed, in parallel. In stage 3, the accumulated sum for $x_1 \cdot y$ is computed and combined with that of $x_0 \cdot y$ to provide the final result  $x \cdot y$.

Our proposed multiplier has a latency of 3 cycles and an amortized throughput of 1 multiplication per cycle. A convenient benefit of this approach is that it's easily re-configurable for different integer sizes and DSP sizes: simply adjust the chunk size of $y$ to match the DSP's bitwidth and change the partition of $x$ to adjust the final multiplication result latency.

\subsection{254-bit modular multiplication unit}
\label{sec:mod-multiplier}

The three-stage 254-bit integer multipliers described in Section~\ref{sec:int-multiplier} are utilized as building blocks to implement a fast pipelined modular multiplication module. This module uses the Barrett multiplication method as shown in Algorithm~\ref{alg:GFp_mult_barrett}. Since there are three large integer multiplications to be performed sequentially, we instantiate three 254-bit integer multipliers, each assigned to one of the three multiplications. Each integer multiplier requires 3 cycles to finish its work, but one extra cycle is needed to transfer the result to the intermediate pipeline registers and forward it to the next integer multiplier. Without this intermediate step, the wires are prone to be too long in the inferred circuit, degrading the clock frequency. Hence, each integer multiplication takes 4 cycles, resulting in a 12-cycle pipeline for the whole modular multiplication. This module has a latency of 12 cycles and an amortized throughput of 1 modular multiplication per cycle.

\subsubsection{Modular multiplication without DSPs} In scenarios where the FPGA board used is too limited and does not offer any DSPs at all, we also present a 254-bit modular multiplication module based on the Russian Peasant algorithm, described by Algorithm~\ref{alg:GFp_mult_peasant}. This algorithm, also referred to as the shift-and-add method, does not use any integer multiplication and hence does not require DSPs at all. The primary disadvantage of this approach is the longer compute time: requiring 254 clock cycles per modular multiplication.

\subsection{Fast $x^7$ modular exponentiation}
\label{sec:fast-exp}

\begin{figure}
    \centering
    \begin{subfigure}{0.55\columnwidth}
    \includegraphics[width=\columnwidth]{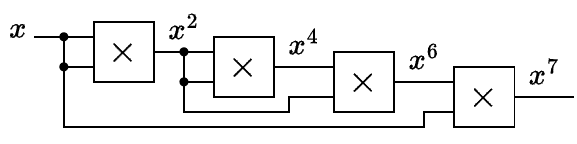}
    \caption{\label{fig:mod-exp-a}}
    \end{subfigure}
    \begin{subfigure}{0.55\columnwidth}
    \includegraphics[width=0.8\columnwidth]{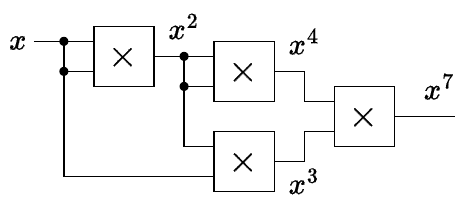}
    \caption{\label{fig:mod-exp-b}}
    \end{subfigure}
    \caption{Computation of $x^7$ modular exponentiation using (a) one modular multiplication unit with a total delay equivalent to four modular multiplications and (b) two modular multiplication units in parallel to obtain a total delay equivalent to three modular multiplications.}
    \vspace{-2mm}
\end{figure}

\begin{figure}
    \begin{subfigure}{\columnwidth}
    \centering
    \includegraphics[width=0.6\columnwidth]{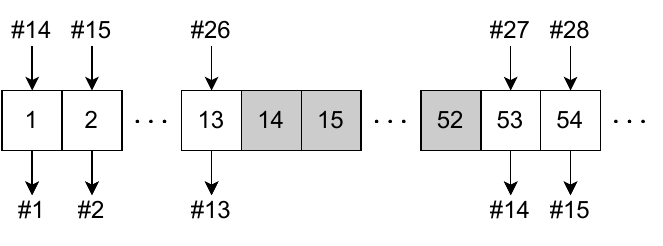}
    \caption{\label{fig:mod-exp-a-pipeline}}
    \end{subfigure}
    \begin{subfigure}{\columnwidth}
    \centering
    \includegraphics[width=0.6\columnwidth]{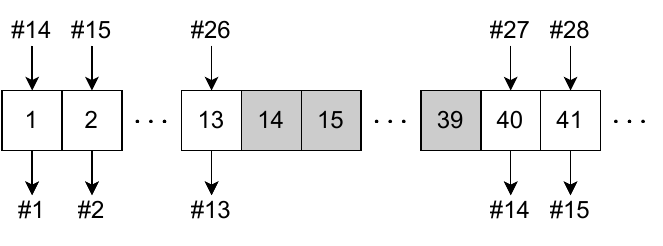}
    \caption{\label{fig:mod-exp-b-pipeline}}
    \end{subfigure}
    \caption{Pipeline scheme for $x^7$ modular exponentiation using (a) one modular multiplication unit with a total delay equivalent to four modular multiplications and (b) two modular multiplication units in parallel to obtain a total delay equivalent to three modular multiplications.}
    \vspace{-5mm}
\end{figure}

The naive method to compute the $x^7$ modular exponentiation consists of six successive modular multiplications by $x$. This approach is inefficient; in practice, one can achieve the same result with four modular multiplications by applying the ``exponentiation by squaring'' technique. This approach is illustrated in Figure~\ref{fig:mod-exp-a}, in which a single modular multiplication block is used to perform the four multiplications in sequence. Naturally, this results in a total delay equivalent to four modular multiplications. Since one modular multiplication takes 12 cycles, and 1 extra cycle is needed for pipeline register transfers as explained in Section~\ref{sec:mod-multiplier}, therefore it would take a total of $(12+1) \times 4 = 52$ cycles to compute $x^7$. The proposed pipeline scheme is shown in Figure~\ref{fig:mod-exp-a-pipeline}. Since there is only one modular multiplication instance present, a pipelined design cannot accept new computation requests in all of its cycles. Therefore, this pipeline only accepts new requests in the first 13 cycles of the whole 52-cycle pipeline. This results in a latency of 52 cycles and an amortized throughput of $13 \mathbin{/} 52 = 0.25$ modular exponentiations per cycle.

However, if a second modular multiplication block can fit on the FPGA board, another optimized design that takes advantage of performing two multiplications in parallel can be used to implement the modular exponentiation with a total delay equivalent to only three modular multiplications, as shown in Figure~\ref{fig:mod-exp-b}. In this case, it would take a total of $(12+1) \times 3 = 39$ cycles to compute $x^7$, but at the cost of increased requirements of DSPs and other logic resources. The pipeline scheme for this case is shown in Figure~\ref{fig:mod-exp-b-pipeline}. Our proposed design is parameterizable and hence supports both approaches to computing $x^7$, including the underlying modular multiplication algorithm (Russian Peasant or Barrett). The full MiMC design that uses this faster alternative to compute $x^7$ is referred to as \texttt{AMZ-2} in this paper.

%% file: sections/5_results.tex
\section{Results}

\subsection{Experimental Setup}
Almost all presented results are from designs developed in a Quartus environment, and synthesized on a Stratix V GT FPGA. While \sys targets resource-constrained devices, we also show that our novel methodology can take advantage of resources to provide even lower latency. We perform tests on a Kintex Ultrascale+ in a Vivado environment to show how \sys scales to larger boards. Alongside this, we show that \sys can be scaled down when necessary, at the cost of latency. We perform tests on a Artix-7 AC701 in a Vivado environment to show how \sys can be utilized with even more constraints on the available hardware.
We measure the CPU runtime of the MiMC hash function on an Intel Gold Xeon 6338 CPU with 1TB of memory --- a very powerful machine. This is to ensure that there is a fair comparison between a high-end, modern CPU and our FPGA implementation to further quantify the value of \sys. We utilize the highly-optimized implementation of MiMC from \cite{githubGitHubTetrationLabarkworksmimc}, built using a state-of-the-art Rust framework called Arkworks \cite{githubArkworks}. Note that experiments were performed with the optimized ``release'' build of this software. All constraints and designs of the MiMC hardware accelerator can be found in the \sys repository.

\subsection{\sys-Powered Designs}

In this section, we outline six end-to-end \sys-powered designs for the MiMC block cipher and hash. These designs fully characterize the reconfigurable parameter space and capabilities of \sys by presenting intermediate and final hardware descriptions that are realizable on low-end, resource-constrained FPGAs. We highlight \texttt{AMZ-1} and \texttt{AMZ-2} as the most efficient realizations of MiMC, and provide \texttt{AMZ-3} as a DSP-free approach to computing MiMC. The designs which are pipelined, can service a batch of multiple computation requests in parallel. Depending on the pipeline architecture, each design has a different batch size that it supports. Naturally, the designs that lack pipelined architecture are incapable of servicing multiple computation requests in parallel, and hence their batch size is effectively $1$.

\textbf{Design \texttt{AMZ-1}} uses the Barrett multiplication method, and hence is powered by the optimized modular multiplication unit described in Section~\ref{sec:mod-multiplier} and the optimized integer multiplier described in Section~\ref{sec:int-multiplier}. For modular exponentiation, it uses a single physical modular multiplication unit as shown in Figure~\ref{fig:mod-exp-a}.

\textbf{Design \texttt{AMZ-1a}} is identical to \texttt{AMZ-1}, except that it does not use the optimized integer multiplier (see Section~\ref{sec:int-multiplier}) in its design of the Barrett multiplication unit. Instead, a simple hardware description of multiplication is written in the form ``\texttt{out <= x*y;}'' and the automatically synthesized 254-bit integer multiplication logic is utilized. In this case, the Barrett multiplication unit has a latency of 3 cycles and a throughput of 1 modular multiplication per cycle.

\textbf{Design \texttt{AMZ-1b}} is identical to \texttt{AMZ-1a}, except that it does not use pipeline architecture.

\textbf{Designs \texttt{AMZ-2}, \texttt{AMZ-2a} and \texttt{AMZ-2b}} are identical to the designs \texttt{AMZ-1}, \texttt{AMZ-1a} and \texttt{AMZ-1b}, except that these use two physical modular multiplication units instead of one, for faster modular exponentiation as shown in Figure~\ref{fig:mod-exp-b}.

\textbf{Design \texttt{AMZ-3}} uses the Russian peasant multiplication method and does not use pipeline architecture. In other words, it cannot service multiple MiMC cipher/hash requests in a batch, and hence the any request must wait until the previous request is finished being serviced. However, it does use two physical modular multiplication units for faster modular exponentiation, as shown in Figure~\ref{fig:mod-exp-b}.

\subsection{\sys Evaluation}

\begin{table*}[t]
    \centering\resizebox{0.95\textwidth}{!}{
    \begin{tabular}{cccccccccccc}
        \toprule
        \multicolumn{2}{c}{\textbf{}} & \multicolumn{5}{c}{\textbf{Resource Utilization}} & \multicolumn{4}{c}{\textbf{Timing}} \\
        \cmidrule(lr){3-7}
        \cmidrule(lr){8-11}
         & Batch &  &  &  &  &  & Total & & Amortized & Amortized & \\
         Design & size & ALMs & LUTs & FFs & BRAM (bits) & DSPs & Cycles & Freq. (MHz) & Latency (\textmu s) & Throughput (ops/s) & Power (W)\\
        \midrule
         \texttt{AMZ-1} & 13 & 20,011 (13\%) & 41,227 (13\%) & 24,623 (4\%) & 1,386 (1\%) & 226 (89\%) & 4,823 & 128.27 & 2.892 & 345,781 & 2.753 \\
         \texttt{AMZ-1a} & 4 & 21,898 (14\%) & 43,025 (14\%) & 6,201 (1\%) & 0 (0\%) & 228 (89\%) & 1,547 & 43.47 & 8.896 & 112,410 & 10.722 \\
         \texttt{AMZ-1b} & 1 & 21,449 (14\%) & 42,120 (14\%) & 4,964 (1\%) & 0 (0\%) & 228 (89\%) & 4,189 & 44.89 & 93.316 & 10,716 & 9.579 \\
         \texttt{AMZ-2} & 13 & 59,765 (38\%) & 120,118 (38\%) & 56,660 (9\%) & 2,772 (1\%) & 218 (86\%) & 3,640 & 125.75 & 2.226 & 449,236 & 3.807 \\
         \texttt{AMZ-2a} & 4 & 68,933 (43\%) & 133,866 (42\%) & 7,389 (2\%) & 0 (0\%) & 224 (88\%) & 1,183 & 39.97 & 7.399 & 135,153 & 11.508 \\
         \texttt{AMZ-2b} & 1 & 68,236 (43\%) & 133,816 (42\%) & 17,687 (3\%) & 0 (0\%) & 224 (88\%) & 3,370 & 44.07 & 76.469 & 13,077 & 12.185 \\
         \texttt{AMZ-3} & 1 & 4,514 (3\%) & 7,391 (3\%) & 5,518 (1\%) & 0 (0\%) & 0 (0\%) & 72,028 & 151.45 & 475.589 & 2,102 & 1.176 \\
         \bottomrule
    \end{tabular}}
    \caption{Comparison of all \sys-powered designs, evaluated on resource utilization, timing, and power.}
    \label{tab:amz_benchmark}
    \vspace{-6mm}
\end{table*}


To evaluate \sys, we focus on evaluating the MiMC block cipher construction itself. The MiMC hash construction is just a ``thin wrapper'' around the block cipher and hence does not create much performance overhead in the final hardware design. Any such hash computation is mostly a sequence of repeated MiMC block cipher invocations --- the length of this sequence is dependent on the length of the message being hashed. The cost of each hash round is dominated by the cost of one block cipher computation. We assume the software running on the CPU will pad the message and embed the bytes into the Galois field GF($p$) as well. Thus, the cost of preparing the message for hashing and the cost of CPU-FPGA communication are out of the scope of this evaluation.

In Table~\ref{tab:amz_benchmark}, we show the performance of various \sys-powered designs on the Stratix V GT board. All the figures are for the final synthesized netlist, as reported by the Quartus tool. We quantify look-up tables (LUTs) as the sum of the combinational adaptive LUTS (ALUTs) used for (a) logic, (b) route-throughs, and (3) memory. We quantify flip-flops (FFs) as the sum of the registers used for (a) design implementation, and (b) routing optimization. The synthesizer was configured to aggressively optimize for performance/speed and use the highest level of fitting effort. The power utilization figures are estimated with all estimator parameters having their default values. The percentages in parentheses show the normalized figures for the FPGA board, rounded up to integer values. Note that the latency and throughput figures in Table~\ref{tab:amz_benchmark} are \textit{amortized} over long time. In other words, the latency shows the cost of a single MiMC block cipher computation over a very long duration of time, assuming all requests are made in full batches in order to saturate the pipeline 100\%.

As can be seen in Table \ref{tab:amz_benchmark}, we are able to achieve low utilization for all FPGA resources, bar DSPs, for all of our designs. Our featured design, \texttt{AMZ-1}, strikes the best balance between area efficiency, power efficiency, and computational throughput: it achieves one of the lowest power consumption levels as well as one of the lowest amortized latency. Although \texttt{AMZ-2} achieves lower latency, it comes at the cost of significantly more synthesized logic in the form of LUTs and adaptive logic modules (ALMs) as well as high a power draw. This is due to the fact that \texttt{AMZ-2} instantiates two modular multiplication hardware units instead of just one. The same relationship can be observed between the \texttt{(a)} and \texttt{(b)} variants of the designs \texttt{AMZ-1} and \texttt{AMZ-2}. E.g. \texttt{AMZ-2a} achieves higher throughput than \texttt{AMZ-1a} but also consumes more logic resources and draws more power. The notable takeaways from both the \texttt{(a)} designs are the effects of our optimized integer multiplier module --- there is a significant degradation of latency, resource efficiency, and power efficiency in the designs \texttt{AMZ-1a} and \texttt{AMZ-2a} as they use the automatic implementation generated by the synthesizer and are missing the applied optimizations present in \texttt{AMZ-1} and \texttt{AMZ-2}. Similarly, there is a massive decrease in performance in both the \texttt{(b)} designs because these designs lack our pipeline architecture, in addition to lacking the optimized integer multiplier unit. We include these adjacent designs to show the impact of \sys's carefully crafted hardware modules for Galois field arithmetic and the MiMC block cipher. Finally, we present the results for the \texttt{AMZ-3} design, which is able to compute the MiMc hash with the lowest power and resource utilization, most notably with zero DSPs, at the significant cost of significantly lower throughput. We highlight that this design is only suitable for devices with a significant lack of computational power.

\begin{table}[htb!]
    \centering
    \resizebox{0.95\columnwidth}{!}{
    \begin{tabular}{cccc}
    \toprule
         \textbf{Device} & \textbf{Frequency (MHz)} & \textbf{Latency (\textmu s)} & \textbf{Power (W)}\\
         \midrule
         CPU & - & 31.093 & 214.5\\
         Artix-7 & 58.82 & 6.307 & 1.2\\
         Stratix V GT & 128.27 & 2.892 & 2.7 \\
         Kintex Ultrascale+ & 156.25 & 2.374 & 3.2\\
         \bottomrule
    \end{tabular}}
    \caption{Comparison of \texttt{AMZ-1} design on different sizes, powers, and speeds devices. \texttt{AMZ-1} is modified to use 16-bit chunks for partial multiplication for Artix-7 and Kintex Ultrascale+ as it better suits the smaller bit-widths of the device's DSPs.}
    \label{tab:cpu_comp}
    \vspace{-12mm}
\end{table}

In Table~\ref{tab:cpu_comp}, we compare the performance of the \texttt{AMZ-1} design on three FPGAs of varying size and power and a powerful CPU. As can be seen, the performance of \texttt{AMZ-1} improves as more expensive hardware is available. The Artix-7, a very low-end and relatively inexpensive FPGA, is still able to achieve 5$\times$ speedup when compared to the CPU. When \texttt{AMZ-1} is implemented on a high-end Kintex Ultrascale+ board, we notice an improvement in latency compared to the Stratix board and an increase in clock frequency and power usage. This is a testament to \sys's ability to take advantage of extra computational resources and power when available, but still achieve practical latency and power consumption when there are heavier constraints on resources. Nevertheless, in all scenarios, \sys enables efficient hardware accelerators on FPGAs of any size that can outperform CPU in latency and power consumption.


%% file: sections/6_conclusion.tex
\section{Conclusion}
In this work, we present \sys, a highly-optimized, parameterized hardware architecture framework for computing the MiMC block cipher and ZK-friendly hash function, a core operation in zero-knowledge proofs, on FPGA. \sys is the first framework to target resource-constrained devices for a ZK-friendly hash function. Through our extensive evaluation, we show that the most efficient MiMC design that can be built with \sys achieves more than 13$\times$ speedup when compared to a state-of-the-art CPU implementation (optimized software written in Rust). Alongside the most efficient design, we present several adjacent designs that take advantage of \sys's parameterizable implementation, such as a resource-optimized, DSP-free MiMC implementation. Our framework introduces a paradigm shift in resource-aware hardware acceleration of ZK-friendly hashes that are necessary to enable zero-knowledge proofs on resource-constrained devices. The open-source nature of our work ensures that developers can easily utilize \sys in their development of custom zero-knowledge applications that require an efficient MiMC implementation. While this work focuses on the important and prominent MiMC ZK-friendly hash function, \sys can be used to accelerate any ZK-friendly hash function that relies on Galois field arithmetic over a large prime field. This work represents an important first step in the direction of practical as well as openly accessible zero-knowledge applications on the edge.

%% file: main.bbl

\begin{thebibliography}{48}


\ifx \showCODEN    \undefined \def \showCODEN     #1{\unskip}     \fi
\ifx \showDOI      \undefined \def \showDOI       #1{#1}\fi
\ifx \showISBNx    \undefined \def \showISBNx     #1{\unskip}     \fi
\ifx \showISBNxiii \undefined \def \showISBNxiii  #1{\unskip}     \fi
\ifx \showISSN     \undefined \def \showISSN      #1{\unskip}     \fi
\ifx \showLCCN     \undefined \def \showLCCN      #1{\unskip}     \fi
\ifx \shownote     \undefined \def \shownote      #1{#1}          \fi
\ifx \showarticletitle \undefined \def \showarticletitle #1{#1}   \fi
\ifx \showURL      \undefined \def \showURL       {\relax}        \fi
\providecommand\bibfield[2]{#2}
\providecommand\bibinfo[2]{#2}
\providecommand\natexlab[1]{#1}
\providecommand\showeprint[2][]{arXiv:#2}

\bibitem[git({[n.\,d.]})]%
        {githubGitHubHarryRethsnarks}
 \bibinfo{year}{[n.\,d.]}\natexlab{}.
\newblock \bibinfo{title}{{G}it{H}ub - {H}arry{R}/ethsnarks: {A} toolkit for viable zk-{S}{N}{A}{R}{K}{S} on {E}thereum, {W}eb, {M}obile and {D}esktop --- github.com}.
\newblock \bibinfo{howpublished}{\url{https://github.com/HarryR/ethsnarks}}.
\newblock
\newblock
\shownote{[Accessed 06-05-2024]}.


\bibitem[ama(2024)]%
        {amazonAmazonInstances}
 \bibinfo{year}{2024}\natexlab{}.
\newblock \bibinfo{title}{{A}mazon {E}{C}2 {F}1 {I}nstances --- aws.amazon.com}.
\newblock \bibinfo{howpublished}{\url{https://aws.amazon.com/ec2/instance-types/f1/}}.
\newblock
\newblock
\shownote{[Accessed 02-05-2024]}.


\bibitem[git(2024a)]%
        {githubArkworks}
 \bibinfo{year}{2024}\natexlab{a}.
\newblock \bibinfo{title}{arkworks - github.com}.
\newblock
\newblock
\newblock
\shownote{[Accessed 06-05-2024]}.


\bibitem[git(2024b)]%
        {githubGitHubBsdevlinfpga_snark_prover}
 \bibinfo{year}{2024}\natexlab{b}.
\newblock \bibinfo{title}{{G}it{H}ub - bsdevlin/fpga\_snark\_prover: {A}n acceleration engine for proving {S}{N}{A}{R}{K}{S} over the bn128 curve, targeted for {A}{W}{S} {F}{P}{G}{A}s --- github.com}.
\newblock \bibinfo{howpublished}{\url{https://github.com/bsdevlin/fpga\_snark\_prover}}.
\newblock
\newblock
\shownote{[Accessed 02-05-2024]}.


\bibitem[git(2024c)]%
        {githubGitHubDatenlordTRIDENT}
 \bibinfo{year}{2024}\natexlab{c}.
\newblock \bibinfo{title}{{G}it{H}ub - datenlord/{T}{R}{I}{D}{E}{N}{T}: {A} {H}ardware {I}mplemented {P}oseidon {H}asher --- github.com}.
\newblock \bibinfo{howpublished}{\url{https://github.com/datenlord/TRIDENT.git}}.
\newblock
\newblock
\shownote{[Accessed 03-05-2024]}.


\bibitem[int(2024)]%
        {intelStratixFPGA}
 \bibinfo{year}{2024}\natexlab{}.
\newblock \bibinfo{title}{{S}tratix® {V} {G}{X} {F}{P}{G}{A} {D}evelopment {K}it --- intel.com}.
\newblock \bibinfo{howpublished}{\url{https://www.intel.com/content/www/us/en/products/details/fpga/development-kits/stratix/v-gx.html}}.
\newblock
\newblock
\shownote{[Accessed 02-05-2024]}.


\bibitem[git(024x)]%
        {githubGitHubZcashFoundationzcashfpga}
 \bibinfo{year}{2024x}\natexlab{}.
\newblock \bibinfo{title}{{G}it{H}ub - {Z}cash{F}oundation/zcash-fpga: {Z}cash {F}{P}{G}{A} acceleration engine --- github.com}.
\newblock \bibinfo{howpublished}{\url{https://github.com/ZcashFoundation/zcash-fpga/tree/master}}.
\newblock
\newblock
\shownote{[Accessed 02-05-2024]}.


\bibitem[Albrecht et~al\mbox{.}(2016)]%
        {albrecht2016mimc}
\bibfield{author}{\bibinfo{person}{Martin Albrecht}, \bibinfo{person}{Lorenzo Grassi}, \bibinfo{person}{Christian Rechberger}, \bibinfo{person}{Arnab Roy}, {and} \bibinfo{person}{Tyge Tiessen}.} \bibinfo{year}{2016}\natexlab{}.
\newblock \showarticletitle{MiMC: Efficient encryption and cryptographic hashing with minimal multiplicative complexity}. In \bibinfo{booktitle}{\emph{International Conference on the Theory and Application of Cryptology and Information Security}}. Springer, \bibinfo{pages}{191--219}.
\newblock


\bibitem[Albrecht et~al\mbox{.}(2019)]%
        {albrecht2019feistel}
\bibfield{author}{\bibinfo{person}{Martin~R Albrecht}, \bibinfo{person}{Lorenzo Grassi}, \bibinfo{person}{L{\'e}o Perrin}, \bibinfo{person}{Sebastian Ramacher}, \bibinfo{person}{Christian Rechberger}, \bibinfo{person}{Dragos Rotaru}, \bibinfo{person}{Arnab Roy}, {and} \bibinfo{person}{Markus Schofnegger}.} \bibinfo{year}{2019}\natexlab{}.
\newblock \showarticletitle{Feistel structures for MPC, and more}. In \bibinfo{booktitle}{\emph{Computer Security--ESORICS 2019: 24th European Symposium on Research in Computer Security, Luxembourg, September 23--27, 2019, Proceedings, Part II 24}}. Springer, \bibinfo{pages}{151--171}.
\newblock


\bibitem[Aly et~al\mbox{.}(2020)]%
        {aly2020design}
\bibfield{author}{\bibinfo{person}{Abdelrahaman Aly}, \bibinfo{person}{Tomer Ashur}, \bibinfo{person}{Eli Ben-Sasson}, \bibinfo{person}{Siemen Dhooghe}, {and} \bibinfo{person}{Alan Szepieniec}.} \bibinfo{year}{2020}\natexlab{}.
\newblock \showarticletitle{Design of symmetric-key primitives for advanced cryptographic protocols}.
\newblock \bibinfo{journal}{\emph{IACR Transactions on Symmetric Cryptology}} (\bibinfo{year}{2020}), \bibinfo{pages}{1--45}.
\newblock


\bibitem[Barreto et~al\mbox{.}(2003)]%
        {barreto2003constructing}
\bibfield{author}{\bibinfo{person}{Paulo~SLM Barreto}, \bibinfo{person}{Ben Lynn}, {and} \bibinfo{person}{Michael Scott}.} \bibinfo{year}{2003}\natexlab{}.
\newblock \showarticletitle{Constructing elliptic curves with prescribed embedding degrees}. In \bibinfo{booktitle}{\emph{Security in Communication Networks: Third International Conference, SCN 2002 Amalfi, Italy, September 11--13, 2002 Revised Papers 3}}. Springer, \bibinfo{pages}{257--267}.
\newblock


\bibitem[Barreto and Naehrig(2005)]%
        {barreto2005pairing}
\bibfield{author}{\bibinfo{person}{Paulo~SLM Barreto} {and} \bibinfo{person}{Michael Naehrig}.} \bibinfo{year}{2005}\natexlab{}.
\newblock \showarticletitle{Pairing-friendly elliptic curves of prime order}. In \bibinfo{booktitle}{\emph{International workshop on selected areas in cryptography}}. Springer, \bibinfo{pages}{319--331}.
\newblock


\bibitem[Barrett(1986)]%
        {barrett1986implementing}
\bibfield{author}{\bibinfo{person}{Paul Barrett}.} \bibinfo{year}{1986}\natexlab{}.
\newblock \showarticletitle{Implementing the Rivest Shamir and Adleman public key encryption algorithm on a standard digital signal processor}. In \bibinfo{booktitle}{\emph{Conference on the Theory and Application of Cryptographic Techniques}}. Springer, \bibinfo{pages}{311--323}.
\newblock


\bibitem[Ben-Sasson et~al\mbox{.}(2018)]%
        {ben2018scalable}
\bibfield{author}{\bibinfo{person}{Eli Ben-Sasson}, \bibinfo{person}{Iddo Bentov}, \bibinfo{person}{Yinon Horesh}, {and} \bibinfo{person}{Michael Riabzev}.} \bibinfo{year}{2018}\natexlab{}.
\newblock \showarticletitle{Scalable, transparent, and post-quantum secure computational integrity}.
\newblock \bibinfo{journal}{\emph{Cryptology ePrint Archive}} (\bibinfo{year}{2018}).
\newblock


\bibitem[Ben-Sasson et~al\mbox{.}(2017)]%
        {ben2017scalable}
\bibfield{author}{\bibinfo{person}{Eli Ben-Sasson}, \bibinfo{person}{Alessandro Chiesa}, \bibinfo{person}{Eran Tromer}, {and} \bibinfo{person}{Madars Virza}.} \bibinfo{year}{2017}\natexlab{}.
\newblock \showarticletitle{Scalable zero knowledge via cycles of elliptic curves}.
\newblock \bibinfo{journal}{\emph{Algorithmica}}  \bibinfo{volume}{79} (\bibinfo{year}{2017}), \bibinfo{pages}{1102--1160}.
\newblock


\bibitem[Ben-Sasson et~al\mbox{.}(2020)]%
        {ben2020stark}
\bibfield{author}{\bibinfo{person}{Eli Ben-Sasson}, \bibinfo{person}{Lior Goldberg}, {and} \bibinfo{person}{David Levit}.} \bibinfo{year}{2020}\natexlab{}.
\newblock \showarticletitle{Stark friendly hash--survey and recommendation}.
\newblock \bibinfo{journal}{\emph{Cryptology ePrint Archive}} (\bibinfo{year}{2020}).
\newblock


\bibitem[Benvenuto(2012)]%
        {benvenuto2012galois}
\bibfield{author}{\bibinfo{person}{Christoforus~Juan Benvenuto}.} \bibinfo{year}{2012}\natexlab{}.
\newblock \showarticletitle{Galois field in cryptography}.
\newblock \bibinfo{journal}{\emph{University of Washington}} \bibinfo{volume}{1}, \bibinfo{number}{1} (\bibinfo{year}{2012}), \bibinfo{pages}{1--11}.
\newblock


\bibitem[Bertoni et~al\mbox{.}(2007)]%
        {bertoni2007sponge}
\bibfield{author}{\bibinfo{person}{Guido Bertoni}, \bibinfo{person}{Joan Daemen}, \bibinfo{person}{Micha{\"e}l Peeters}, {and} \bibinfo{person}{Gilles Van~Assche}.} \bibinfo{year}{2007}\natexlab{}.
\newblock \showarticletitle{Sponge functions}. In \bibinfo{booktitle}{\emph{ECRYPT hash workshop}}, Vol.~\bibinfo{volume}{2007}.
\newblock


\bibitem[Boo et~al\mbox{.}(2021)]%
        {boo2021litezkp}
\bibfield{author}{\bibinfo{person}{EunSeong Boo}, \bibinfo{person}{Joongheon Kim}, {and} \bibinfo{person}{JeongGil Ko}.} \bibinfo{year}{2021}\natexlab{}.
\newblock \showarticletitle{LiteZKP: Lightening zero-knowledge proof-based blockchains for IoT and edge platforms}.
\newblock \bibinfo{journal}{\emph{IEEE Systems Journal}} \bibinfo{volume}{16}, \bibinfo{number}{1} (\bibinfo{year}{2021}), \bibinfo{pages}{112--123}.
\newblock


\bibitem[Bowe et~al\mbox{.}(2020)]%
        {bowe2020zexe}
\bibfield{author}{\bibinfo{person}{Sean Bowe}, \bibinfo{person}{Alessandro Chiesa}, \bibinfo{person}{Matthew Green}, \bibinfo{person}{Ian Miers}, \bibinfo{person}{Pratyush Mishra}, {and} \bibinfo{person}{Howard Wu}.} \bibinfo{year}{2020}\natexlab{}.
\newblock \showarticletitle{Zexe: Enabling decentralized private computation}. In \bibinfo{booktitle}{\emph{2020 IEEE Symposium on Security and Privacy (SP)}}. IEEE, \bibinfo{pages}{947--964}.
\newblock


\bibitem[Bsdevlin({[n.\,d.]})]%
        {Bsdevlin}
\bibfield{author}{\bibinfo{person}{Bsdevlin}.} \bibinfo{year}{[n.\,d.]}\natexlab{}.
\newblock \bibinfo{title}{Fpga\_snark\_prover/fpga\_snark\_prover/kernel/readme.md at master · bsdevlin/FPGA\_SNARK\_PROVER}.
\newblock
\newblock
\urldef\tempurl%
\url{https://github.com/bsdevlin/fpga\_snark\_prover/blob/master/fpga\_snark\_prover/kernel/README.md}
\showURL{%
\tempurl}


\bibitem[Chen et~al\mbox{.}(2022)]%
        {chen2022review}
\bibfield{author}{\bibinfo{person}{Thomas Chen}, \bibinfo{person}{Hui Lu}, \bibinfo{person}{Teeramet Kunpittaya}, {and} \bibinfo{person}{Alan Luo}.} \bibinfo{year}{2022}\natexlab{}.
\newblock \showarticletitle{A review of zk-snarks}.
\newblock \bibinfo{journal}{\emph{arXiv preprint arXiv:2202.06877}} (\bibinfo{year}{2022}).
\newblock


\bibitem[Chen et~al\mbox{.}(2023)]%
        {chen2023survey}
\bibfield{author}{\bibinfo{person}{Zhigang Chen}, \bibinfo{person}{Yuting Jiang}, \bibinfo{person}{Xinxia Song}, {and} \bibinfo{person}{Liqun Chen}.} \bibinfo{year}{2023}\natexlab{}.
\newblock \showarticletitle{A survey on zero-knowledge authentication for internet of things}.
\newblock \bibinfo{journal}{\emph{Electronics}} \bibinfo{volume}{12}, \bibinfo{number}{5} (\bibinfo{year}{2023}), \bibinfo{pages}{1145}.
\newblock


\bibitem[Coron et~al\mbox{.}(2005)]%
        {coron2005merkle}
\bibfield{author}{\bibinfo{person}{Jean-S{\'e}bastien Coron}, \bibinfo{person}{Yevgeniy Dodis}, \bibinfo{person}{C{\'e}cile Malinaud}, {and} \bibinfo{person}{Prashant Puniya}.} \bibinfo{year}{2005}\natexlab{}.
\newblock \showarticletitle{Merkle-Damg{\aa}rd revisited: How to construct a hash function}. In \bibinfo{booktitle}{\emph{Advances in Cryptology--CRYPTO 2005: 25th Annual International Cryptology Conference, Santa Barbara, California, USA, August 14-18, 2005. Proceedings 25}}. Springer, \bibinfo{pages}{430--448}.
\newblock


\bibitem[Gaba et~al\mbox{.}(2022)]%
        {gaba2022zero}
\bibfield{author}{\bibinfo{person}{Gurjot~Singh Gaba}, \bibinfo{person}{Mustapha Hedabou}, \bibinfo{person}{Pardeep Kumar}, \bibinfo{person}{An Braeken}, \bibinfo{person}{Madhusanka Liyanage}, {and} \bibinfo{person}{Mamoun Alazab}.} \bibinfo{year}{2022}\natexlab{}.
\newblock \showarticletitle{Zero knowledge proofs based authenticated key agreement protocol for sustainable healthcare}.
\newblock \bibinfo{journal}{\emph{Sustainable Cities and Society}}  \bibinfo{volume}{80} (\bibinfo{year}{2022}), \bibinfo{pages}{103766}.
\newblock


\bibitem[Ghodsi et~al\mbox{.}(2023)]%
        {ghodsi2023zprobe}
\bibfield{author}{\bibinfo{person}{Zahra Ghodsi}, \bibinfo{person}{Mojan Javaheripi}, \bibinfo{person}{Nojan Sheybani}, \bibinfo{person}{Xinqiao Zhang}, \bibinfo{person}{Ke Huang}, {and} \bibinfo{person}{Farinaz Koushanfar}.} \bibinfo{year}{2023}\natexlab{}.
\newblock \showarticletitle{zPROBE: Zero peek robustness checks for federated learning}. In \bibinfo{booktitle}{\emph{Proceedings of the IEEE/CVF International Conference on Computer Vision}}. \bibinfo{pages}{4860--4870}.
\newblock


\bibitem[Grassi et~al\mbox{.}(2021)]%
        {grassi2021poseidon}
\bibfield{author}{\bibinfo{person}{Lorenzo Grassi}, \bibinfo{person}{Dmitry Khovratovich}, \bibinfo{person}{Christian Rechberger}, \bibinfo{person}{Arnab Roy}, {and} \bibinfo{person}{Markus Schofnegger}.} \bibinfo{year}{2021}\natexlab{}.
\newblock \showarticletitle{Poseidon: A new hash function for $\{$Zero-Knowledge$\}$ proof systems}. In \bibinfo{booktitle}{\emph{30th USENIX Security Symposium (USENIX Security 21)}}. \bibinfo{pages}{519--535}.
\newblock


\bibitem[Grassi et~al\mbox{.}(2023)]%
        {grassi2023poseidon2}
\bibfield{author}{\bibinfo{person}{Lorenzo Grassi}, \bibinfo{person}{Dmitry Khovratovich}, {and} \bibinfo{person}{Markus Schofnegger}.} \bibinfo{year}{2023}\natexlab{}.
\newblock \showarticletitle{Poseidon2: A faster version of the poseidon hash function}. In \bibinfo{booktitle}{\emph{International Conference on Cryptology in Africa}}. Springer, \bibinfo{pages}{177--203}.
\newblock


\bibitem[Groth(2016)]%
        {groth2016size}
\bibfield{author}{\bibinfo{person}{Jens Groth}.} \bibinfo{year}{2016}\natexlab{}.
\newblock \showarticletitle{On the size of pairing-based non-interactive arguments}. In \bibinfo{booktitle}{\emph{Advances in Cryptology--EUROCRYPT 2016: 35th Annual International Conference on the Theory and Applications of Cryptographic Techniques, Vienna, Austria, May 8-12, 2016, Proceedings, Part II 35}}. Springer, \bibinfo{pages}{305--326}.
\newblock


\bibitem[Ingonyama(2022)]%
        {Ingonyama_2022}
\bibfield{author}{\bibinfo{person}{Ingonyama}.} \bibinfo{year}{2022}\natexlab{}.
\newblock \bibinfo{title}{Systemization of knowledge: ZK-friendly hash functions}.
\newblock
\newblock
\urldef\tempurl%
\url{https://medium.com/@ingonyama/system-of-knowledge-zk-friendly-hash-functions-ab825616c9f1}
\showURL{%
\tempurl}


\bibitem[Khovratovich and Vladimirov(2019)]%
        {khovratovich2019tornado}
\bibfield{author}{\bibinfo{person}{Dmitry Khovratovich} {and} \bibinfo{person}{Mikhail Vladimirov}.} \bibinfo{year}{2019}\natexlab{}.
\newblock \showarticletitle{Tornado Privacy Solution. Cryptographic Review. Version 1.1}.
\newblock \bibinfo{journal}{\emph{ABDK Consulting, November}}  \bibinfo{volume}{29} (\bibinfo{year}{2019}).
\newblock


\bibitem[Koc et~al\mbox{.}(1996)]%
        {koc1996analyzing}
\bibfield{author}{\bibinfo{person}{C~Kaya Koc}, \bibinfo{person}{Tolga Acar}, {and} \bibinfo{person}{Burton~S Kaliski}.} \bibinfo{year}{1996}\natexlab{}.
\newblock \showarticletitle{Analyzing and comparing Montgomery multiplication algorithms}.
\newblock \bibinfo{journal}{\emph{IEEE micro}} \bibinfo{volume}{16}, \bibinfo{number}{3} (\bibinfo{year}{1996}), \bibinfo{pages}{26--33}.
\newblock


\bibitem[Langhammer and Pasca(2021)]%
        {langhammer2021efficient}
\bibfield{author}{\bibinfo{person}{Martin Langhammer} {and} \bibinfo{person}{Bogdan Pasca}.} \bibinfo{year}{2021}\natexlab{}.
\newblock \showarticletitle{Efficient FPGA modular multiplication implementation}. In \bibinfo{booktitle}{\emph{The 2021 ACM/SIGDA International Symposium on Field-Programmable Gate Arrays}}. \bibinfo{pages}{217--223}.
\newblock


\bibitem[Liu and Ning(2011)]%
        {liu2011zero}
\bibfield{author}{\bibinfo{person}{Hong Liu} {and} \bibinfo{person}{Huansheng Ning}.} \bibinfo{year}{2011}\natexlab{}.
\newblock \showarticletitle{Zero-knowledge authentication protocol based on alternative mode in RFID systems}.
\newblock \bibinfo{journal}{\emph{IEEE Sensors Journal}} \bibinfo{volume}{11}, \bibinfo{number}{12} (\bibinfo{year}{2011}), \bibinfo{pages}{3235--3245}.
\newblock


\bibitem[Liu et~al\mbox{.}(2021)]%
        {liu2021zkcnn}
\bibfield{author}{\bibinfo{person}{Tianyi Liu}, \bibinfo{person}{Xiang Xie}, {and} \bibinfo{person}{Yupeng Zhang}.} \bibinfo{year}{2021}\natexlab{}.
\newblock \showarticletitle{Zkcnn: Zero knowledge proofs for convolutional neural network predictions and accuracy}. In \bibinfo{booktitle}{\emph{Proceedings of the 2021 ACM SIGSAC Conference on Computer and Communications Security}}. \bibinfo{pages}{2968--2985}.
\newblock


\bibitem[Lu et~al\mbox{.}(2008)]%
        {lu2008pseudo}
\bibfield{author}{\bibinfo{person}{Li Lu}, \bibinfo{person}{Jinsong Han}, \bibinfo{person}{Yunhao Liu}, \bibinfo{person}{Lei Hu}, \bibinfo{person}{Jin-Peng Huai}, \bibinfo{person}{Lionel Ni}, {and} \bibinfo{person}{Jian Ma}.} \bibinfo{year}{2008}\natexlab{}.
\newblock \showarticletitle{Pseudo trust: Zero-knowledge authentication in anonymous P2Ps}.
\newblock \bibinfo{journal}{\emph{IEEE Transactions on Parallel and Distributed Systems}} \bibinfo{volume}{19}, \bibinfo{number}{10} (\bibinfo{year}{2008}), \bibinfo{pages}{1325--1337}.
\newblock


\bibitem[Ma et~al\mbox{.}(2023)]%
        {ma2023gzkp}
\bibfield{author}{\bibinfo{person}{Weiliang Ma}, \bibinfo{person}{Qian Xiong}, \bibinfo{person}{Xuanhua Shi}, \bibinfo{person}{Xiaosong Ma}, \bibinfo{person}{Hai Jin}, \bibinfo{person}{Haozhao Kuang}, \bibinfo{person}{Mingyu Gao}, \bibinfo{person}{Ye Zhang}, \bibinfo{person}{Haichen Shen}, {and} \bibinfo{person}{Weifang Hu}.} \bibinfo{year}{2023}\natexlab{}.
\newblock \showarticletitle{Gzkp: A gpu accelerated zero-knowledge proof system}. In \bibinfo{booktitle}{\emph{Proceedings of the 28th ACM International Conference on Architectural Support for Programming Languages and Operating Systems, Volume 2}}. \bibinfo{pages}{340--353}.
\newblock


\bibitem[Menezes et~al\mbox{.}(2018)]%
        {menezes2018handbook}
\bibfield{author}{\bibinfo{person}{Alfred~J Menezes}, \bibinfo{person}{Paul~C Van~Oorschot}, {and} \bibinfo{person}{Scott~A Vanstone}.} \bibinfo{year}{2018}\natexlab{}.
\newblock \bibinfo{booktitle}{\emph{Handbook of applied cryptography}}.
\newblock \bibinfo{publisher}{CRC press}.
\newblock


\bibitem[Ni and Zhu(2023)]%
        {ni2023enabling}
\bibfield{author}{\bibinfo{person}{Ning Ni} {and} \bibinfo{person}{Yongxin Zhu}.} \bibinfo{year}{2023}\natexlab{}.
\newblock \showarticletitle{Enabling zero knowledge proof by accelerating zk-SNARK kernels on GPU}.
\newblock \bibinfo{journal}{\emph{J. Parallel and Distrib. Comput.}}  \bibinfo{volume}{173} (\bibinfo{year}{2023}), \bibinfo{pages}{20--31}.
\newblock


\bibitem[Nitulescu(2020)]%
        {nitulescu2020zk}
\bibfield{author}{\bibinfo{person}{Anca Nitulescu}.} \bibinfo{year}{2020}\natexlab{}.
\newblock \showarticletitle{zk-SNARKs: a gentle introduction}.
\newblock \bibinfo{journal}{\emph{Ecole Normale Superieure}} (\bibinfo{year}{2020}).
\newblock


\bibitem[Pan and Meher(2013)]%
        {pan2013bit}
\bibfield{author}{\bibinfo{person}{Yu Pan} {and} \bibinfo{person}{Pramod~Kumar Meher}.} \bibinfo{year}{2013}\natexlab{}.
\newblock \showarticletitle{Bit-level optimization of adder-trees for multiple constant multiplications for efficient FIR filter implementation}.
\newblock \bibinfo{journal}{\emph{IEEE Transactions on Circuits and Systems I: Regular Papers}} \bibinfo{volume}{61}, \bibinfo{number}{2} (\bibinfo{year}{2013}), \bibinfo{pages}{455--462}.
\newblock


\bibitem[Sharma et~al\mbox{.}(2020)]%
        {sharma2020blockchain}
\bibfield{author}{\bibinfo{person}{Bhavye Sharma}, \bibinfo{person}{Raju Halder}, {and} \bibinfo{person}{Jawar Singh}.} \bibinfo{year}{2020}\natexlab{}.
\newblock \showarticletitle{Blockchain-based interoperable healthcare using zero-knowledge proofs and proxy re-encryption}. In \bibinfo{booktitle}{\emph{2020 International Conference on COMmunication Systems \& NETworkS (COMSNETS)}}. IEEE, \bibinfo{pages}{1--6}.
\newblock


\bibitem[Sklavos and Koufopavlou(2003)]%
        {sklavos2003hardware}
\bibfield{author}{\bibinfo{person}{Nicolas Sklavos} {and} \bibinfo{person}{Odysseas Koufopavlou}.} \bibinfo{year}{2003}\natexlab{}.
\newblock \showarticletitle{On the hardware implementations of the SHA-2 (256, 384, 512) hash functions}. In \bibinfo{booktitle}{\emph{Proceedings of the 2003 International Symposium on Circuits and Systems, 2003. ISCAS'03.}}, Vol.~\bibinfo{volume}{5}. IEEE, \bibinfo{pages}{V--V}.
\newblock


\bibitem[Tetration-Lab(2024)]%
        {githubGitHubTetrationLabarkworksmimc}
\bibfield{author}{\bibinfo{person}{Tetration-Lab}.} \bibinfo{year}{2024}\natexlab{}.
\newblock \bibinfo{title}{GitHub - Tetration-Lab/arkworks-mimc: Arkworks implementation of cryptographic hash function MiMC}.
\newblock
\newblock
\newblock
\shownote{[Accessed 06-05-2024]}.


\bibitem[Weng et~al\mbox{.}(2021)]%
        {weng2021mystique}
\bibfield{author}{\bibinfo{person}{Chenkai Weng}, \bibinfo{person}{Kang Yang}, \bibinfo{person}{Xiang Xie}, \bibinfo{person}{Jonathan Katz}, {and} \bibinfo{person}{Xiao Wang}.} \bibinfo{year}{2021}\natexlab{}.
\newblock \showarticletitle{Mystique: Efficient conversions for Zero-Knowledge proofs with applications to machine learning}. In \bibinfo{booktitle}{\emph{30th USENIX Security Symposium (USENIX Security 21)}}. \bibinfo{pages}{501--518}.
\newblock


\bibitem[Wu et~al\mbox{.}(2020)]%
        {wu2020blockchain}
\bibfield{author}{\bibinfo{person}{Wei Wu}, \bibinfo{person}{Erwu Liu}, \bibinfo{person}{Xinglin Gong}, {and} \bibinfo{person}{Rui Wang}.} \bibinfo{year}{2020}\natexlab{}.
\newblock \showarticletitle{Blockchain based zero-knowledge proof of location in iot}. In \bibinfo{booktitle}{\emph{ICC 2020-2020 IEEE International Conference on Communications (ICC)}}. IEEE, \bibinfo{pages}{1--7}.
\newblock


\bibitem[Xavier(2022)]%
        {xavier2022pipemsm}
\bibfield{author}{\bibinfo{person}{Charles~F Xavier}.} \bibinfo{year}{2022}\natexlab{}.
\newblock \showarticletitle{Pipemsm: Hardware acceleration for multi-scalar multiplication}.
\newblock \bibinfo{journal}{\emph{Cryptology ePrint Archive}} (\bibinfo{year}{2022}).
\newblock


\bibitem[Zhang et~al\mbox{.}(2021)]%
        {zhang2021pipezk}
\bibfield{author}{\bibinfo{person}{Ye Zhang}, \bibinfo{person}{Shuo Wang}, \bibinfo{person}{Xian Zhang}, \bibinfo{person}{Jiangbin Dong}, \bibinfo{person}{Xingzhong Mao}, \bibinfo{person}{Fan Long}, \bibinfo{person}{Cong Wang}, \bibinfo{person}{Dong Zhou}, \bibinfo{person}{Mingyu Gao}, {and} \bibinfo{person}{Guangyu Sun}.} \bibinfo{year}{2021}\natexlab{}.
\newblock \showarticletitle{Pipezk: Accelerating zero-knowledge proof with a pipelined architecture}. In \bibinfo{booktitle}{\emph{2021 ACM/IEEE 48th Annual International Symposium on Computer Architecture (ISCA)}}. IEEE, \bibinfo{pages}{416--428}.
\newblock


\end{thebibliography}
